\documentclass[12pt]{article}
\usepackage{amssymb,amsmath}
\usepackage{openbib}
\usepackage{amsthm}
\usepackage[dvips]{graphicx}
\theoremstyle{plain}
\newtheorem{lem}{Lemma}
\newtheorem{prop}{Proposition}
\newtheorem{cor}{Corollary}
\newtheorem{theor}{Theorem}

\newtheorem{conjecture}{Conjecture}

\def\ola{\overleftarrow}
\def\ora{\overrightarrow}
\def\cf{{\it cf. }}

\def\ie{i.e.\ }

\def\R{{\mathbb{R}}}      % real numbers
\def\id{{id}}              % alternate identity operator
\def\wt{\widetilde}       % wide tilde
\def\A{\mathcal{A}}       % potnerial 1 form
\def\S{\mathcal{S}}       % Schwartz space
\def\L{\mathcal{L}}
\def\M{\mathcal{M}}

\def\R{\mathbb{R}}
\def\Z{\mathbb{Z}}

\def\tr{\mbox {\rm tr} }

\def\itn#1{{\rm{(#1)}}}    % roman italic enumeration
\def\q{\hat q}
\def\p{\hat p}
\def\f{\hat f}
\def\h{\hbar}

\def\w{\wedge}
\def\pr{\partial}
\def\ih{\frac i \h }
\def\L2{L^2(\R^n)}
\def\pf{\noindent{\it{Proof}}\,.}

\def\d{{\delta^j_s}}
\def\f{\hat f}
\def\g{\hat g}
\def\x{\hat x}
\def\F{{F_{sj}}}
\def\h{\hbar}
\def\w{\wedge}

\def\theequation{\arabic{section}.\arabic{equation}}
\begin{document}

\begin{titlepage}

\font\title=cmbx12 \centerline{{}} \vspace{1cm}
 \centerline{{\title
Quantum Magnetic Algebra and Magnetic Curvature}}
 \vspace{15 mm}
 \centerline{M. V. Karasev} \vskip 8pt
\centerline{{\sl Department of Applied Mathematics}} \centerline{{\sl Moscow
Institute of Electronics and Mathematics}} \centerline{{\sl Moscow 109028,
Russia}}

\vspace{10mm} \centerline{T. A. Osborn } \vskip 8pt \centerline{{\sl Department
of Physics and Astronomy}} \centerline{{\sl University of Manitoba}}
\centerline{{\sl Winnipeg, MB, Canada, R3T 2N2 }}

\vspace{10 mm}

\begin{abstract}

The symplectic geometry of the phase space associated with a charged particle
is determined by the addition of the Faraday 2-form to the standard $dp\w dq$
structure on $\R^{2n}$. In this paper we describe the corresponding algebra of
Weyl-symmetrized functions in operators $\hat q$, $\hat p$ satisfying nonlinear
commutation relations. The multiplication in this algebra generates an
associative $*$ product of functions on the phase space. This $*$ product is
given by an integral kernel whose phase is the symplectic area of a
groupoid-consistent membrane. A symplectic phase space connection with
non-trivial curvature is extracted from the magnetic reflections associated
with the Stratonovich quantizer.  Zero and constant curvature cases are
considered as examples. The quantization with both static and time dependent
electromagnetic fields is obtained. The expansion of the $*$ product by the
deformation parameter $\h$, written in the covariant form, is compared with the
known deformation quantization formulas.\vfill

\end{abstract}
\end{titlepage}

\section{Introduction}

 An associative noncommutative multiplication of functions on phase space,
corresponding to the Poisson structure, is called a {\it quantization}. On
general phase spaces, \ie on symplectic manifolds having a symplectic
connection, there is now a well developed scheme of deformation quantization
\cite{BBF78,Raw91, OMY91, BF96,Kon97,BNW98}. This fundamental theory is
formulated in symplectic terms, but is based on formal asymptotic expansions
and generally does not have an operator representation in Hilbert space.

Only a few examples are known where the quantization is {\it{perfect}}, that
is:\vskip 2mm

\indent - it is exact, rather than expressed via asymptotic series;

\indent - it has an operator representation in a Hilbert space (corresponding
to Schr\"odinger
 \indent \ \ quantum mechanics); and

\indent - it is given explicitly and purely in symplectic terms. \vskip 2mm

\noindent The first two conditions from this list are realized in certain
examples of the strict deformation quantization \cite{Rief89,Biel01} and in the
tangential groupoid quantization \cite{Lan93,Car99}, but the last (geometrical)
criterion is not fulfilled.

The only known perfect examples are related to the phase space $T^*\R^n =
\R^{2n}$  with canonical symplectic structure and the trivial connection, or to
cylinder-type spaces, where the coordinates are subject to periodicity
conditions \cite{FAB72}, or to a generic symplectic form  but with a flat (zero
curvature) symplectic connection \cite{BT90,GB92}. We do not refer here to the
homogeneous K\"ahlerian spaces where the quantization by coherent states could
be considered as perfect, but involves symplectic areas in a complexified phase
space.

The example of perfect quantization which we construct in the present paper is
concerned with non-zero and non-constant phase space curvature. The formalism
allows one to represent, in a manifestly gauge invariant and covariant manner,
the dynamics of a charged particle in an electromagnetic field realized in
terms of a quantum phase space.

There are two ways to introduce the magnetic coupling into quantum mechanics.
The first is based on the interpretation of the magnetic potential as a
connection form in the $U(1)$-bundle over the configuration space and
subsequently considers the corresponding modification of the dynamical
(Schr\"odinger, Klein--Gordon, etc.) equations. The second approach
incorporates the idea of modifying the usual symplectic form $dp\wedge dq$ on
phase space by adding the Faraday 2-form and then to quantize this new
symplectic space, and in particular, to represent functions on this space by
operators. The present paper employs this second method.

We consider the phase space $\R^{2n}=\R^n_q\oplus\R^n_p$ with the following
`magnetic' symplectic form
\begin{equation}\label{I.1}
\omega = dp\wedge dq + \frac 12 F(q)\, dq\wedge dq\,.
\end{equation}
 The coordinates $p$ have the physical interpretation of the gauge invariant (kinetic) momenta
and $F$ is a skew tensor on the configuration space $\R^{n}_q$
representing the magnetic portion of the electromagnetic field \cf
\cite{BS82}. The closedness of the form (\ref{I.1}) is equivalent
to the homogeneous Maxwell equation for the Faraday tensor $F$.
The charge coupling constant and the speed of light are all set
equal to 1.

The non-degenerate symplectic form (\ref{I.1}) is a simple modification of the
canonical form $dp\wedge dq$, but nevertheless the appearance of a generic
tensor $F$ makes the structure of the quantum phase space rather nontrivial. In
general, we assume that the components of the tensor $F(q)$ are nonlinear
functions of $q$, but note that the linear and constant cases still provide
interesting physical examples.

Our procedure for constructing the quantum phase space is the following. First,
we quantize the Poisson brackets related to the form $\omega$ and immediately
obtain the commutation relations between the quantum coordinates
\begin{equation}\label {I.2}
[\q^j,\q^s]=0, \qquad [\q^j,\p_s]=i\h\d, \qquad [\p_j,\p_s]=i\h\F(\q)\,.
\end{equation}
The nonlinearity of $F$ in the momentum commutation relations means that Lie
algebra techniques are not applicable here.  Furthermore, as we shall see, it
is precisely this nonlinearity  that is responsible for the appearance of
quantum phase space curvature.

It is easy to represent relations (\ref{I.2}) in terms of self-adjoint
operators on $L^2(\R^n)$ and then to construct the Weyl--symmetrized functions
of those operators, specifically
\begin{equation}\label{I.2b}
\f= f(\q,\p) = f \bigg( \frac {\stackrel {1} {\hat q} + \stackrel {3} {\hat q}
} 2 ,\, \stackrel {2} {\hat p} \bigg)\,.
\end{equation}
The over numbering of the operators indicates the order in which they act on a
target wave function as in \cite{Mas73}. Thus, we have a linear mapping $f
\rightarrow\f$. Considering this mapping as an operator-valued linear
functional we can represent it in the integral form
\begin{equation}\label{I.3}
\f = \int f(x)\, \Delta(x)\, dx\,. \end{equation} Here $x=(q,p)$ denote phase
space points. In this way we shall obtain a family of operators (quantizers)
$\Delta(x)$ acting in the same Hilbert space where the representation of
algebra (\ref{I.2}) is given. De-quantization, the inverse map to (\ref{I.3}),
is also constructed by the quantizer. We shall see that for a suitable class of
operators,
\begin{equation}\label{I.3a}
f(x)= (2\pi\h)^n \,\tr\, \big(\f\, \Delta(x)\big)\,.
\end{equation}
We say that $f$ is the (magnetic) {\it symbol} image of the operator $\f$.

The family $\Delta$ in the case of zero magnetic tensor was first introduced in
\cite{STR57} and has since been intensively studied for algebras with linear
commutation relations in \cite{Car99,GB92,VGB89}.

For general nonlinear tensors $F$ in (\ref{I.2}) the quantizer $\Delta$ is
still well-defined and possesses the following basic properties:

\indent (i) the elements of $\Delta$ are linearly independent, invertible and
resolve the identity;

\indent (ii) the linear envelope of elements of $\Delta$ form an algebra.

The `structure constants'  of this algebra generate a non-commutative $*$
product of functions on phase space,
\begin{equation}\label{I.4}
\widehat {f*g} = \f \g\,.
\end{equation} We call $*$ a {\it magnetic product}. This product satisfies
the correspondence principle:
\begin{equation*}  f*g =fg - \frac
{i\h}2 \{f,g\} + O(\h^2)\quad\mbox{as}\quad \h \rightarrow 0
\end{equation*} (as is usual in the deformation quantization scheme), where
$\{\cdot,\cdot\}$ denotes the Poisson bracket related to the symplectic form
(\ref{I.1}) and $fg$ is the commutative product of functions.

Our main goal is to interpret this magnetic product geometrically, and to
demonstrate how the quantizer generates a phase space connection.

First, we shall see that the quantizer $\Delta(z)$ generates a symplectic
transformation $\sigma_z:x\rightarrow x'$ in $\R^{2n}$. This transformation is
given by the Fock--type formula \cite{Fock}:
\begin{equation}\label{I.5}
\Delta(z)^{-1}\x \Delta(z) = \x'\,.
\end{equation}
Here $\x=(\q,\p)$ is the set of generators appearing in (\ref{I.2}), and
$\x'=(\q',\p')$ is a new set (with the same commutation relations). The symbol
image of (\ref{I.5}) defines $\sigma_z$.

For each $z\in\R^{2n}$ the mapping $\sigma_z:\R^{2n}\rightarrow\R^{2n}$
preserves the symplectic form $\omega$, has the fixed point $z=\sigma_z(z)$,
and is an involution: ${\sigma_z}^2=\id$. We call $\sigma=\{\sigma_z\}$ a
family of {\it {magnetic reflections}}. Using these reflections one can realize
the symplectic groupoid multiplication rule corresponding to relations
(\ref{I.2}) \cite{KO1,KM91}. Then  for each triplet of points $z,y,x$ we can
construct a membrane $\Sigma(z,y,x)$ in $\R^{2n}$ whose boundary is consistent
with the groupoid structure; namely, a boundary consisting of three linked
$\sigma$-reflective curves with mid-points $z,y,x$.

Based on our previous results \cite{KO1,KO02} we shall obtain the following
formula for the magnetic non-commutative product:
\begin{equation}\label{I.7}
(f * g )(z) = \frac 1{(\pi\h)^{2n}}\int\!\!\int\exp\Big\{ \frac i \hbar
\int_{\Sigma(z,y,x)}\omega \Big\} f(y) g(x)\,dy \,d x\,.
\end{equation}

So, we see that the membrane WKB phase of the $*$ product integral kernel,
conjectured in \cite{AW94} for symmetric symplectic manifolds, is realized in
the magnetic phase space exactly, without the need for a WKB expansion.

The product (\ref{I.7}) is strict (not formal). Its asymptotic expansion as
$\hbar \rightarrow 0$ can be written in the bi-differential covariant form:
\begin{equation} \label{I.8}
f*g =fg - \frac {i\h}2 f\langle   \ola\nabla\Psi\ora\nabla\rangle g - \frac
{\h^2}{8} f\langle \ola\nabla\Psi\ora\nabla\rangle^2 g + O(\h^3)\,.
\end{equation}
Here $ \Psi =[\begin{array}{cc}0&-I\\I&F\end{array}]$ is the Poisson tensor
corresponding to the symplectic structure $\omega$. The covariant derivative
$\nabla$ acts either on the left multiplier or the right multiplier as
indicated by the arrows but does not act on the argument of $\Psi$. The
derivative above corresponds to a connection on the phase space $\R^{2n}$
defined by the following Christoffel symbols,
\begin{equation} \label{I.9}
\Gamma^j_{sl}(x)  = -\frac 12 \frac {\pr^2\sigma_z(x)^j}{\pr x^s \pr x^l}{\bigg
|}_{z=x}\,.
\end{equation}
We call this a {\it magnetic connection}. It is symplectic: $\nabla \omega =
0$. We emphasize that this phase space connection is generated by the tensor
$F$ given on configuration space $\R^n_q$, but $\Gamma$ is not a Riemannian
type connection.

All the higher terms $O(\hbar^k)$ in (\ref{I.8}) contain the Poisson bracket
contribution $f\langle \ola\nabla\Psi \ora\nabla\rangle^k g$ plus additional
terms generated by the curvature of $\Gamma$, so that
\begin{equation}\label{I.10}
f*g = f\exp\Big\{-{\frac {i\h}2} \langle\ola \nabla \Psi(x) \ora \nabla\rangle\
+ \frac{i\h^3}{48} R^{ijkl}(\ora\nabla_i\ora\nabla_j\ora\nabla_k\ola\nabla_l -
\ola\nabla_i\ola\nabla_j\ola\nabla_k\ora\nabla_l) + O(\h^4)\Big\} g \,,
\end{equation}
where $R$ is the curvature tensor with all indices raised by the Poisson
structure.

If the tensor $F$ is quadratic in Euclidean coordinates on $\R^n_q$, then the
curvature tensor $R$ on the phase space $\R^{2n}=T^*\R^n$ is constant but
non-zero. This case provides a rather interesting example of a symmetric
symplectic space (in the terminology of \cite{BCG95}). Here the mappings
$\sigma$ possess an additional property
\begin{equation}\label{I.11}
\sigma_y\sigma_z\sigma_y(x) = \sigma_{\sigma_y(z)}(x)\,, \qquad \forall \,\,
x,y,z
\end{equation}
and the connection (\ref{I.9}) is recognized as the Cartan--Loos canonical
connection \cite{Loo69} corresponding to the family of $\sigma$-symmetries.

In the Riemannian setting, Cartan \cite{CART26} called this class of spaces
``remarkable". In our magnetic framework this symmetric structure does not
belong to Riemannian geometry but still is remarkable. The commutation
relations (\ref{I.2}) in this case look quadratic
\begin{equation}\label {I.13}
[\q^j,\q^s]=0\,, \qquad [\q^j,\p_s]=i\h\,\delta^j_s\,, \qquad [\p_j,\p_s]=
{i\h}\, F_{js,kl}\, \q^k \q^l\,.
\end{equation}
The integral formula (\ref{I.7}) for the associative product corresponding to
this quadratic relations employs membranes $\Sigma(z,y,x)$ bounded by just
geodesics of the Cartan--Loos connection. In the bi-differential formula
(\ref{I.10}) the $O(\h^4)$ remainder  vanishes in this case. \bigskip

So, in summary, the new features developed in the paper are:\smallskip

\indent - construction of a phase space connection and curvature generated by a
generic (electro) magnetic tensor\,,

\indent -  geometric groupoid interpretation of membrane areas in the integral
formula for the associative product corresponding to the non-linear commutation
relations (\ref{I.2})\,,

\indent - realization of the symmetric symplectic structure related to
quadratic brackets (\ref{I.13}) and its explicit quantization by means of
geodesically bounded membranes or by the curvature generated bi-differential
exponent.

\bigskip

This paper is organized as follows. Section 2 describes representations of the
quantizer and the construction of the reflection map $\sigma_x$. The magnetic
$*$ product and its groupoid aspects are discussed in Section 3. The magnetic
connection is found in Section 4 and the curvature features of the $\h$
deformation expansion the $*$ product are presented there.  The role of the
electric field is clarified in Section 5. The next two sections treat the zero
curvature and constant curvature cases.

%The preprint version of this paper in found in the arXiv: phys. ???

\section{Quantizer and magnetic reflections}
\setcounter{equation}{0}\setcounter{lem}{0}

First we make a general remark about the operator calculations made below. All
of them are simple direct constructions and all the formulas are obtained
explicitly, although from the view point of functional analysis the
presentation of results often looks formal.  But actually the suppressed
functional analysis details are standard (about this see the remarks at the end
of Section 3) and of limited usefulness in clarifying the new objects and
results coming out of the calculus. Of course, it is known that the use of
formal methods in non-commutative analysis (even for algebras with the simplest
Heisenberg commutation relations) can lead to errors. A list of problems
demonstrating the `dangerous' areas where the formal analysis gives incorrect
results is found in the book \cite{KM91}, Appendix 1. However, the derivations
below are far from these sensitive analytical areas.

In this section we review the definition of the magnetic quantizer and
introduce the associated reflective structure.

The irreducible representation of commutation relations (\ref{I.2}) in the
Hilbert space $L^2(\R^n)$ is given by the operators
\begin{equation}\label{II.1}
\q^j:\psi(q')\mapsto q'^j \psi(q')\,,\qquad \p_j: \psi(q')\mapsto -i\h \frac
{\pr \psi(q')}{\pr q'^j} - A_j(q')\psi(q')\,.
\end{equation}
Here $q'$ is running over $\R^n$, and  $A_j$ are components of the 1-from
$\A=A_j(q')\,dq'^j$ which is a primitive of the Faraday 2-form $\frac 1 2
F_{jk}(q')\, dq'^k\w dq'^j$, namely
\begin{equation}\label{II.2}
\frac {\pr A_j(q')}{\pr q'^k} - \frac {\pr A_k(q')}{\pr q'^j} = F_{jk}(q')\,.
\end{equation}

The operators (\ref{II.1}) are well defined on a dense domain in $L^2(\R^n)$
and essentially self-adjoint. Thus one can consider Weyl symmetrized functions
of these operators following the general definitions in
\cite{ANDER69,ANDER70,KM91}. In detail, we take smooth and rapidly decaying
functions $f=f(x)$, introduce their Fourier transform $\tilde f$ and obtain
operators in $\L2$ via
\begin{equation}\label{II.3}
\f = \int \tilde f(\eta) \exp\{\frac i \h \eta\cdot \x\}\, d\eta\,,\quad
\x=(\q,\p)\,.
\end{equation}
It is easy to see that
\begin{equation}\label{II.4}
\exp\{\frac i \h \eta\cdot \x\} = \exp\{\frac i {2\h} \eta_q\cdot \q\}
\exp\{\frac i {\h} \eta_p\cdot \p\}\exp\{\frac i {2\h} \eta_q\cdot \q\}\,,
\end{equation}
where $\eta_q$ and $\eta_p$ are just the components of the vector $\eta\in
\R^n_q\oplus\R^n_p$.

From the factorization (\ref{II.4}) we observe that formula (\ref{I.2b})
follows. Also, from the definition of momentum operators (\ref{II.1}) we have
\begin{eqnarray*}
&\exp{\Big \{{\displaystyle{\ih}} \eta_p\cdot \p\Big\}}
 &= \exp\Big\{\eta_p\cdot \frac \pr{\pr q'} - \ih \eta_p \cdot A(q')\Big\}\\
& &= \exp\Big\{\eta_p\cdot \stackrel {2} {\frac \pr{\pr q'}} -\ih \int_0^1
\eta_p\cdot A((1-\tau) \stackrel {3} {q'} + \tau\stackrel {1} {q'})\,d\tau\Big\}\\
& &= \exp \Big\{ -\ih \int_0^1 \eta_p\cdot A((1-\tau)\,q' +
\tau(q'+\eta_p))\,d\tau\Big\} \exp\Big\{\eta_p \cdot \frac \pr {\pr q'}\Big\}
\end{eqnarray*}
So, on any function $\psi\in \L2$ this operator exponential acts as follows:
\begin{eqnarray*}
&\exp\Big\{{\displaystyle{\ih}} \eta_p\cdot \p\Big\}\psi(q') &=
\exp\Big\{-\ih\int_0^1\eta_p\cdot A(q' +\tau \eta_p) \,d\tau\Big\}
\psi(q'+\eta_p)\qquad\\& &= \exp\Big\{ -\ih \int_{q'}^{q'+\eta_p} \A \Big\}
\psi(q'+\eta_p) \,,
\end{eqnarray*}
where the integral of the 1-form $\A$ is taken along the straight-line segment
in $\R^n$ connecting $q'$ to $q'+\eta_p$.

Thus formula (\ref{II.4}) implies,
\begin{equation}
\exp\big\{ \ih \eta \cdot \x \big\} \psi(q') = \exp\Big\{ \frac i{2\h} \eta_q
\cdot \eta_p -\ih \int_{q'}^{q'+\eta_p}\A + \ih \eta_q \cdot q' \Big\}
\psi(q'+\eta_p)\,.
\end{equation}
Taking the inverse Fourier transform of (\ref{II.3}) and representing the
operators $\f$ in the form (\ref{I.3}) we obtain the quantizer acting on the
function $\psi$,
\begin{equation}\label{II.5}
\Delta(q,p)\,\psi(q') = {\frac 1{(\pi\h)^{n}}}  \exp\Big\{ \frac {2i}{\h}
p\cdot (q'-q) +\ih \int_{2q-q'}^{q'} \A \Big\} \psi(2q-q')\,.
\end{equation}
Observe that the point $2q-q'$ in the rightmost $\psi$ is just the original
$q'$ reflected through the point $q$ with respect to the Euclidean structure on
$\R^n$. The integral kernel statement equivalent to (\ref{II.5}) is
\begin{equation}\label{II.5a}
\langle q'|\Delta(q,p)|q''\rangle = (2\pi\h)^{-n} \delta\Big(\frac {q'+q''}2
-q\Big) \exp\Big\{\ih p \cdot (q'-q'') + \ih\int_{q''}^{q'} \A\Big\}\,.
\end{equation}
The delta function forces the value of the midpoint $(q'+q'')/2$ to be $q$. The
representations (\ref{II.5}) and (\ref{II.5a}) are similar to those used by
Stratonovich \cite{Str56} in defining the gauge invariant Wigner transform.

Using the above representations it is easy to verify the following properties
of the family of operators $\Delta(x)\,\, (x=(q,p)\in\R^{2n})$ acting in the
Hilbert space $\L2$.

\begin{lem}\label{lem2.1}
\begin{eqnarray*}
&\itn{i}&\int \Delta(x)\, d\,x = I\,;\qquad \qquad \qquad
\quad \quad \quad\,\,\\
&\itn{ii} &\Delta(x)^\dagger= \Delta(x)\,, \qquad   \Delta(x)^2 = \frac
1{(\pi\h)^{2n}} I\,; \vphantom\sum \qquad \qquad \qquad
\qquad \quad \quad \quad\,\\
&\itn{iii} &\tr\, \Delta(x_1) \Delta(x_2) = (2\pi\h)^{-n}\delta(x_1-x_2)\,,
\qquad \tr\,\Delta(x) = (2\pi\h)^{-n} \,. \quad \phantom{\bigg )}
\end{eqnarray*}
\end{lem}

The trace used above is understood in the distributional sense, so that the
generalized functions $\tr \Delta(x)$ and  $\tr\big( \Delta(y)\,\Delta(x)\big)$
are naturally defined as distributions on $\R^{2n}$ and $\R^{2n}\times\R^{2n}$
respectively, that is the operator valued functions $\Delta$ are first
integrated in the Bochner sense with test functions and after that the trace
operation is applied.  The operator identities and equations above and
throughout the text are considered in the sense of the strong topology on a
dense domain and then extended (where possible) to the whole Hilbert space
$L^2(\R^n)$.

The equation $\itn{i}$ in Lemma 1 says that $\{\Delta(x)\}$ is a resolution of
the identity; as an example of (\ref{I.3}) it states that the unit symbol
$f(x)=1$ is quantized to the identity operator, $\f=I$. The property $\itn{ii}$
shows that $\Delta(x)$ are bounded self-adjoint operators in $\L2$ with norm
$(\pi\h)^{-n}$; modulo a rescaling, $\Delta(x)$ is unitary operator. The first
part of $\itn{iii}$ in combination with (\ref{I.3}) establishes that the
de-quantization map $\f \rightarrow f $ is given by the trace identity
(\ref{I.3a}) and, as a consequence, that the symbol of $\Delta(x)$ is the delta
function $\delta_x$.

We now consider the construction of the reflection map induced by $\Delta(x)$.
Given (\ref{II.5}) the intertwining identities readily follow
\begin{equation}\label{II.7}\begin{array}{ll}
\q^j\,\Delta(q,p)&\!\!\!= \Delta(q,p)\,(2q-\q)^j\,, \phantom{\bigg )} \\
\p_j\,\Delta(q,p)&\!\!\!= \Delta(q,p)\,(2p-\p - \hat \alpha_q)_j\,\,, \quad

\end{array}j=1,..,n.\end{equation} Here $\alpha_q$ is the following vector function
\begin{equation}\label{II.9a}
\alpha_q(q')= A(q') + A(2q-q') - \frac \pr{\pr q'}\bigg(\int_{2q-q'}^{q'}
\A\bigg)\,.
\end{equation}
A simple calculation shows one can restate this in a gauge invariant fashion as
the average of the magnetic tensor $F$,
\begin{equation}\label{II.10n}
\alpha_q (q') = \int^1_{-1} F(q+\mu(q'-q)) (q'-q)\,\mu\, d\mu\,.
\end{equation}

Left multiply (\ref{II.7}) by $\Delta(x)^{-1}$ and thereby obtain
$x'\rightarrow\sigma_x(x')$ as
\begin{equation}\label{II.9}
\sigma_x(x')= 2x-x' - \left( \begin{array}{c} 0 \\ \alpha_{x_q}(x'_q)
\end{array} \right)\,,
\end{equation}
where $x_q$ and $x'_q$ denote the $q$-components of the phase space points $x$
and $x'$.

\begin{lem} \label{lem2.2} The family of mappings $\{\sigma_x\}$ in the
magnetic phase space $\prec\! \R^{2n},\omega\!\succ$ possesses the following
properties

\itn{i} $\sigma_x(x)=x\,, \quad \forall \,x \phantom{\bigg )}$\qquad \itn{ii}
$\sigma_x^2 = \id $

\itn{iii} $\sigma_x$ is symplectic, \ie preserves the magnetic {\rm 2}-form
$\omega$.
\end{lem}
\noindent{\it Proof}. Part \itn{i} follows from $\alpha_q(q)=0$, and \itn{ii}
from $\alpha_q(2q-q')=\alpha_q(q')$. Property \itn{iii} is a result of the fact
that the operators $\x' =\hat {\sigma_z}$  and $\x$ satisfy the same
commutation relations (\ref{I.2}). So the magnetic 2-form $\omega$ must be
invariant under the change of variables $x'\rightarrow \sigma_x(x')$. $\square$

Property $\itn{ii}$ is the symbol equivalent of the involution identity,
$\Delta(x)^2 =(\pi\h)^{-2n} I$. Also note that the pullback invariance,
$\sigma_x^*\omega=\omega$, means that the vector function $\alpha_q$ satisfies
the tensor identity,
\begin{equation}
\frac {\pr \alpha_q(q')_j}{\pr q^{\prime k}} - \frac {\pr \alpha_q(q')_k}{\pr
q^{\prime j}} = F_{jk}(q') - F_{jk}(2q-q')\,,
\end{equation}
which easily follows from (\ref{II.9a}).

The presence of reflective transformations on phase space allows one to
identify a useful family of curves. A continuous, piecewise differentiable
function $x:[-t,t]\mapsto \R^{2n}$  is called a {\it magnetic reflective curve}
with midpoint $x(0)$ if
\begin{equation*}
\sigma_{x(0)}(x(\tau)) = x(-\tau)\,, \quad  \tau \in [-t,t]\,.
\end{equation*}

 Clearly a reflective curve has a central symmetry about its midpoint. These
curves will be used to define the boundary of the symplectic area of
$\Sigma(x,y,z)$ in Lemma 4 below.  In the case where $F=0$, the family of
reflective curves admit straight lines having endpoints $x(-t)$ and $x(t)$ and
$\sigma_x$ becomes the Grossmann--Royer transformation \cite{Gro76,Roy77}. The
reflective curves are the natural generalization of the midpoint/chord
construct introduced by Berezin, Berry and Marinov \cite{FAB72,Ber77,Mar79}.
This generalization was used in \cite{AW94} under the different name: symmetric
curves.

\section{Magnetic multiplication}
\setcounter{equation}{0}

Now we present formulas for the magnetic $*$ product. This product has two
distinct representations. The first is given by a Berezin type integral formula
whose phase involves a three sided symplectic area. The second is a left, right
regular representation expressed in terms of pseudo-differential operators. We
show how the family of magnetic reflections $\sigma$ interrelates these
different representations and how it is associated with the groupoid core of
the $*$ product.

Observe that the linear envelope of quantizers form an algebra. From
(\ref{I.3}) and (\ref{I.3a}) one has,
\begin{equation}\label{III.1} \Delta(y)\Delta(x) = \int\! K(z,y,x) \Delta(z)\, dz\,.
\end{equation}
The complex functions $K$ are the symbols of the operators $\Delta(y)\Delta(x)$
and are regarded as the `structure constants' of this algebra.

    From Lemma\,\ref{lem2.1}\,(iii) it readily follows that
\begin{equation}\label{III.2}
K(z,y,x) = (2\pi\h)^n\, \tr\,\big(\Delta(z)\Delta(y)\Delta(x)\big) ={\frac
1{(\pi\h)^{2n}}} \exp\Big\{{\frac i \hbar} \int_{M(z,y,x)}\omega \Big\}\,,
\end{equation}
where $M(z,y,x)$ is a straight-line triangle having $z,y,x$ as  midpoints of
its sides. The presence of the `midpoint' delta functions in the quantizer
kernel (\ref{II.5a}) collapses all the integrals in the trace giving the simple
exponential result seen above. The symplectic membrane formula (\ref{III.2}) in
the magnetic phase space was first obtained (by another way) in \cite{KO1}. A
non-symplectic version of the three magnetic quantizer trace can be found in
\cite{DMO92}.

The kernel $K$ in (\ref{III.2}) gives us the integral representation of the
magnetic product (\ref{I.4}),
\begin{equation}\label{III.3}
(f*g)(z) = \int\!\!\int K(z,y,x) f(y) g(x)\,dy \,d x\,.
\end{equation}
The continuous function $K(z,y,x)$ is invariant under any cyclic permutation of
arguments; under the permutation of any pair of arguments, $K\rightarrow
\overline K$; it is the constant $(\pi\h)^{-2n}$ if any pair of arguments are
the same. Also note that the trace operation in (\ref{III.2}) is responsible
for the gauge invariance of the $*$ product. Although $\Delta$ has a $U(1)$
gauge dependence, clearly $M,\,\omega$ and $K$ are invariant.

It turns out that the boundary of the membrane $M(z,y,x)$ in (\ref{III.2}) may
be deformed in many different ways while leaving the kernel function $K(z,y,x)$
unchanged.  Among these deformations is a representation stated in terms of the
$\sigma$ reflective curves which incorporates the groupoid properties of the
magnetic product.

To see this, note that the magnetic product can be written
\begin{equation}\label{III.6}
f*g =f(L)g = g(R)f\,,
\end{equation}
where $L=(L_q,L_p)$ and $R=(R_q,R_p)$ are Weyl-symmetrized sets of
pseudo-differential operators on the phase space $\R^{2n}$. Since the map
$f\rightarrow\f$ is Weyl ordered, $R=\overline L$. The associativity of $*$
implies $[L,R]=0$.

 In \cite{KO1} it was established that
\begin{equation}\label{III.7}
\begin{array}{ll}
L_q =q +\frac12{i\hbar} \partial_p\,,\quad
& R_q =q -\frac12{i\hbar} \partial_p\,,\\[2ex]
L_p =p -\frac12{i\hbar} \partial_q-A(L_q,R_q)\,,\quad & R_p =p +\frac12{i\hbar}
\partial_q-A(R_q,L_q)\,.
\end{array}
\end{equation} The vector-function $A$ is the two-point Valatin
potential \cite{Val54} (also referred to as the Schwinger-Fock, or radial gauge
potential in the literature), obeying
\begin{eqnarray}\label{III.8}
\frac {\pr A(q',q'')_j}{\pr q^{\prime k}}
-\frac {\pr A(q',q'')_k}{\pr q^{\prime j}} &=& F_{jk}(q')\,,\\
\label{III.9}(q'-q'')^jA(q',q'')_j &=& 0\,.
\end{eqnarray}
So, with respect to its first argument, the potential $A(q',q'')$ represents a
primitive of the Faraday 2-form, and satisfies the radial gauge condition
(\ref{III.9}). Taken together, equations (\ref{III.8}) and (\ref{III.9})
uniquely determine the potential $A(q',q'')$ giving the explicit formula
\begin{equation}\label{III.10}
A(q',q'') = \int_0^1 F((\tau q' +(1-\tau)q'')(q'-q'')\,\tau\, d\tau\,.
\end{equation} The Valatin potential is the $\tau$-weighted average of the magnetic force on
a unit charge moving with velocity $q'-q''$ from $q''$ to $q'$.

The function $\alpha_q(q')$, defined by (\ref{II.9a}) and used in our
construction (\ref{II.9}) of the magnetic reflections, is related to the
Valatin potential via
\begin{equation}\label{III.11}
\alpha_q(q') = A(q',2q-q') + A(2q-q',q')\,.
\end{equation}
From this equality and from the explicit formulas (\ref{III.7}) it is
straightforward to determine the interrelationship between the $L,R$ operators
and the reflections $\sigma$. This statement requires the introduction of an
extended phase space $ T^*\R^{2n}$.

\begin{lem}\label{lem3} Definition {\rm(\ref{II.9})} of the
magnetic reflection is equivalent to the following operator identities
\begin{equation}\label{III.12}
\sigma_x(R) = L\,.
\end{equation}
Here $L$ and $R$ are the left and right operators {\rm(\ref{III.6})} of the
regular representation of the magnetic algebra. If $L$ and $R$ are represented
by the symbols
\begin{equation}\label{III.13}
L=l(x,-i\h\ \pr/{\pr x})\,, \qquad R=r(x,-i\h \pr /{\pr x})\,,
\end{equation}
then {\rm(\ref{III.12})} is equivalent to
\begin{equation}\label{III.14}
\sigma_x(r(x,\eta)) = l(x,\eta)\,, \qquad \eta\in T^*_x\R^{2n}\,,
\end{equation}
where
\begin{equation}\label{III.12a}
\begin{array}{ll}
l_q(x,\eta) =x_q - \eta_p/2, \quad & l_p(x,\eta) =x_p+ \eta_q/2-A(l_q,r_q),
\\[2ex]
r_q(x,\eta) =x_q + \eta_p/2, \quad & r_p(x,\eta) =x_p-\eta_q/2-A(r_q,l_q).
\end{array}
\end{equation}
\end{lem}

So the left and right vector functions $l,r$ are defined on the space
$T^*\R^{2n}$. Since the transformation $(x,\eta)\rightarrow (l,r)$ has a unique
inverse either $(x,\eta)$ or $(l,r)$ may be used to represent points $m \in
T^*\R^{2n}\approx \R^{2n}\times\R^{2n}$.

  Recall that groupoid multiplication is defined on the extended space
in the following way. Two points $m_2,\,m_1\in T^*\R^{2n}$ are multiplicable
iff $r(m_2)=l(m_1)$, and in this case, their product is $m=m_2\circ m_1$ where
$l(m)=l(m_2)$ and $r(m)=r(m_1)$. The set of units $e$ consists of the points
where $r(m)=l(m)$ or $m=(x,y)$ with $y=0$. The groupoid product $\circ$ is
noncommutative, associative and has inverse $(l,r)^{-1}=(r,l)$. The
transformations $ l:T^*\R^{2n}\rightarrow \R^{2n}\,,\,\,
r:T^*\R^{2n}\rightarrow\R^{2n}\,, $ are left and right (target and source)
mappings of the groupoid structure on $T^*\R^{2n}$ which corresponds to the
symplectic form $\omega$ on $\R^{2n}$ (see the general theory of symplectic
groupoids in \cite{KM91}). In view of (\ref{III.14}), the magnetic reflections
$\sigma_x$ relate the left and right images $l(m)$ and $r(m)$ in the symplectic
groupoid to each other via the central point $x=x(m)$.

A way to visualize how this groupoid structure can be used to construct the
symplectic area phase for the $*$ product is the following. Given the three
points $x_3,x_2,x_1\in\R^{2n}$, solve the equation
\begin{equation}\label{III.12b}
m_3\circ m_2 \circ m_1 = e\,, \end{equation} subject to the central conditions
$x(m_i)=x_i=(q_i,p_i)\,,\, i=1,2,3$. It is easy to see that this problem has a
unique solution. The $q$ projected image of (\ref{III.12b}) is just the
triangle $\delta(q_3,q_2,q_1)$ in $\R_q^n$ defined by its midpoints
$(q_3,q_2,q_1)$. The endpoints of sides of this triangle are the $l_q,r_q$
values appearing the the Valatin potential, $A(l_q,r_q)$. Employing identities
(\ref{III.12a}) fixes the three complementary endpoints $l_p,r_p$. Now consider
a three-sided membrane $\Sigma(x_3,x_2,x_1)$ in $\R^{2n}$. Each side of its
boundary is characterized by a triplet of points $[r_i,x_i,l_i]$ and some
reflective curve that passes through these points. In the same was as in
\cite{KO1} one can check that
\begin{equation}\label{III.15}
\int_{M(x_3,x_2,x_1)} \omega = \int_{\Sigma(x_3,x_2,x_1)} \omega\,.
\end{equation}
This gives us a groupoid consistent boundary for the the $*$ product membrane.
We note the allowed $\Sigma(x_3,x_2,x_1)$ boundary is non-unique or `floppy' in
character. Given the three of sets of points $[l_i,x_i,r_i]$ satisfying the
multiplicable property $l_i=r_{i+1}$ there are many reflective curves that are
consistent with this data. Nevertheless the value $\int_{\Sigma(x_3,x_2,x_1)}
\omega$ is the same for every allowed reflective curve boundary. As a result of
(\ref{III.2}) and (\ref{III.15}) one has the groupoid compatible form of $K$.

\begin{lem}
\begin{equation}\label{III.16}
K(z,y,x) = \exp\Big\{ \frac i \hbar \int_{\Sigma(z,y,x)}\omega \Big\}\,.
\end{equation}
\end{lem}

Now using this formula for the kernel function $K$ we obtain the associative
non-commutative multiplication of functions on phase space by means of
(\ref{I.7}).

In order to specify the set of functions which is closed with respect to the
operation (\ref{I.7}) one needs to require smoothness and growth estimates on
the magnetic field. We say that all the derivatives of the field tensor $F$
have {\it polynomial growth at infinity} if for some $N< \infty$ there are
estimates
\begin{equation}\label{est1}
|D_q^{\bf s} F_{jk}(q)|<C({\bf s)}\,(1+|q|)^N\,,\quad C({\bf s})<\infty\,,
\quad {\bf s}\in \Z_{+}^n\,.
\end{equation}
The growth power $N$ is independent of ${\bf s}$.

\begin{theor}\label{thm1} Let all the derivatives of the field tensor $F$ have polynomial
growth at infinity. Then the Schwartz space  $\S(\R^{2n})$ of all rapidly
decreasing functions is closed with respect to the magnetic product
{\rm(\ref{I.7})}. This associative algebra has the irreducible representation
$f\rightarrow\f$ {\rm(\ref{I.3})} in the Hilbert space $L^2(\R^n)$ so that
relation {\rm(\ref{I.4})} holds.
\end{theor}

The closure property is proved by using a representation of $f*g$ based on the
Fourier transformed symbols $\tilde f,\tilde g$ and suitable integration by
parts manipulations of this representation. Property (\ref{I.4}) is a
consequence of (\ref{III.1}). The irreducibility follows from the fact that the
set of generators $\q,\p$ in (\ref{II.1}) is irreducible.

The algebra $\S(\R^{2n})$ can be  extended in order to include, say,
polynomials in the generators $\q,\p$. But for this one has to place much
stronger restrictions on the $F$ tensor growth.  If $F$ has compact support one
can certainly extend the algebra $\S(\R^{2n})$ to the $\S^\infty(\R^{2n})$
consisting of smooth functions whose derivatives have polynomial growth at
infinity. In this case the function $1\in\S^\infty(\R^{2n})$ represents the
unity element: $f*1=1*f=f$.

We note that, in view of Lemma 1(iii), the magnetic Weyl  correspondence $f
\leftrightarrow \f$ is a unitary isomorphism from $L^2(\R^{2n})$ to the space
of Hilbert--Schmidt operators in $L^2(\R^n)$. In this paper, we do not
undertake a full investigation of the spaces of operators and symbols which
realize the correspondence $f \leftrightarrow \f$. This is a separate technical
(and often not simple) question which has been extensively studied in the
pseudo-differential operator literature
\cite{GB92,Mas73,KM91,Foll89,Hor79,KN78}. The reader can consider all formulas
as formally algebraic or, depending on the formula, assume an appropriate
simple symbol class such as polynomials, smooth rapidly decreasing functions,
etc.

\section{Magnetic connection and $*$ product expansion}
\setcounter{equation}{0}

Let us now consider magnetic  multiplication (\ref{I.7}) as a one-parameter
family of products depending on $\h$. Assume the smooth symbols $f$ and $g$ are
$\h$ independent.

    All the coefficients of the $\h\rightarrow 0$ expansion
\begin{equation}\label{IV.1}
f*g = fg + \sum_{k \geqslant 1} \frac 1{k!} \left(-\frac {i\h}2\right)^k
\,c_k(f,g)
\end{equation} were described in \cite{KO1} in terms of partial derivatives of the functions $f,g$
and of the magnetic tensor $F$ (see also \cite{Mul99}).

Let us recall the structure of the first three coefficients $c_k(f,g)$. Of
course, the leading term is the Poisson bracket corresponding to $\omega$,
\begin{equation}\label{IV.2}
c_1(f,g) = \{f,g\}=  f\langle\ola D\Psi\ora D\rangle g\,.
\end{equation}
The next two terms are
\begin{equation}\label{IV.3}
\begin{array}{rl}
c_2({\cdot},\cdot) &= \langle \ola D \Psi\ora D\rangle ^2 + \frac 2 3 \pr_s
F_{kl}(\ola \pr^k \ora \pr^l \ola \pr^s - \ola \pr^k \ora \pr^l \ora
\pr^s)\,,\\[2ex]
c_3(\cdot,\cdot) &= \langle \ola D \Psi\ora D\rangle^3 + 2 \langle \ola D
\Psi\ora D\rangle\pr_s F_{kl}(\ola\pr^k \ora\pr^l \ola\pr^s - \ola\pr^k
\ora\pr^l \ora\pr^s)\\[2ex]
&\qquad +\,\, \pr^2_{sr} F_{kl}(\ola\pr^k \ora\pr^l \ola\pr^s \ola\pr^r  +
\ola\pr^k \ora\pr^l\ora\pr^s \ora\pr^r)\,.
\end{array}
\end{equation}

We indicate by $D=(\pr/{\pr q},\pr/{\pr p})$ the total derivative with respect
to all the phase space variables and denote by $\pr^k = \pr /{\pr p}_k$  the
derivative by the momentum component. The arrows indicate on which multiplier
(left or right) the derivatives act. None of the derivatives act on the tensor
components of $\Psi$ or $F$. In particular
\begin{eqnarray*}
f(x)\langle \ola D\, \Psi(x)\ora D\rangle^N g(x) &\equiv& (D_{i_1}\cdots
D_{i_N}f)(x)\Psi^{i_1j_1}(x)\cdots \Psi^{i_Nj_N}(x) (D_{j_1}\cdots
D_{j_N}g)(x)\,,\\
\pr^m_{s_1\dots s_m} F_{kl}(q) &\equiv& {\frac {\pr^m}{\pr q^{s_1} \cdots \pr
q^{s_m}} }F_{kl}(q)\,.
\end{eqnarray*}

The disadvantage of formulas like (\ref{IV.3}) is that they are not in
covariant form. In order to have a covariant expression one has to introduce a
suitable connection on $\R^{2n}$ and replace the partial derivative $D$ by a
covariant derivative $\nabla$. Thus the question at this stage is the
following.  What is the connection that is consistent with the magnetic
quantization? Of course, the space $\R^{2n}$ is equipped with the trivial
Euclidean connection. However, this Euclidean connection knows nothing about
the magnetic 2-form $\omega$ and is not a symplectic connection.

    But there is a natural symplectic connection induced by reflections $\sigma$.

\begin{prop}\label{prop1}
The family of magnetic reflections $\{\sigma_x\}$ defines a symplectic
connection with respect to $\omega$ via the Christoffel symbols
{\rm(\ref{I.9})}. The components $\Gamma^i_{jk}$ vanish unless $j,k\leq n$ and
$i>n$. For $i,j,k \leq n$, the non-trivial portion of $\Gamma$ at the point
$x=(q,p)\in\R^{2n}$ is the following
\begin{equation*}
 \Gamma^{i+n}_{jk} \equiv \widetilde \Gamma_{ijk} =\frac 12 \frac
 {\pr^2\alpha_{q'}(q)_i}{\pr q^j \pr q^k}{\bigg |}_{q'=q}\,.
\end{equation*}
The explicit formula in terms of the magnetic tensor is
\begin{equation}\label{IV.4c}
  \widetilde \Gamma_{ijk}= \frac 13 (\pr_kF_{ij} +  \pr_jF_{ik})\,.
\end{equation}
 This
connection has the curvature tensor
\begin{equation*}
R^i_{jkl} =  \frac\pr{\pr x^k} \Gamma^i_{jl} -\frac\pr{\pr x^l} \Gamma^i_{jk}
\end{equation*}
with the following nonzero components{\rm:}
\begin{equation}\label{RT1}
R^{i+n}_{jkl}\equiv \wt R_{ijkl} =\frac 13 \pr^2_{ij}F_{kl}\,.
\end{equation}
\end{prop}

\noindent{\it Proof}\,. From (\ref{I.9}), since the explicit form (\ref{II.9})
of $\sigma_x$ is known in terms of the Valatin potentials (see (\ref{II.10n},
\ref{III.11})), the transformation of $\Gamma$ under an arbitrary
diffeomorphism $x\rightarrow \tilde x = \tilde x(x)$ is easily calculated shows
that $\Gamma$ is a connection on $\R^{2n}$, \cf \cite{KO02}.  The equality for
the curvature results from the commutativity, $\Gamma^i_{sk}\Gamma^s_{jl}-
\Gamma^i_{sl}\Gamma^s_{jk}=0$.

Finally, consider the symplectic nature of this connection. Write the
connection in block matrix form
\begin{equation}\label{IV.6n}
\Gamma^{i}_{j k}=\Gamma[k]^i_j\,,\qquad  \Gamma[k] = \left(\begin{array}{cc} 0 & 0 \\
\widetilde \Gamma[k] & 0\end{array}\right)\,.
\end{equation}
In this notation the covariant derivative of $\omega$ takes the form
\begin{equation*}\label{IV.6}
\nabla_k\omega_{ij} = \big(\pr_k \omega - \Gamma[k]^T \omega - \omega
\Gamma[k]\big)_{ij}
\end{equation*}
One readily finds that \begin{equation*} \nabla_k\omega =
\left(\begin{array}{cc}
\pr_k F -  \wt \Gamma[k] + \wt \Gamma[k]^T& 0 \\
0 & 0\end{array}\right)\,.
\end{equation*} All the terms in the matrix above are individually zero
if $k > n$.  For $k \leq n$ the upper left block is
\begin{equation*}
\pr_k F_{ij} - \wt \Gamma[k]_{ij} + \wt \Gamma[k]_{ji} = \pr_k F_{ij} - \frac
13 (\pr_k F_{ij} + \pr_j F_{ik}) + \frac 13 (\pr_k F_{ji} + \pr_i F_{jk}) =
0\,.
\end{equation*} The last equality is a consequence of the closeness of the
form (\ref{I.1}). $\square$

If in $\widetilde \Gamma_{ijk}$ the factor $1/3$ is replaced by any other
number then $\Gamma$ ceases to be a symplectic connection. Also, observe that
$\Gamma$ is a function of $q$, but not of $p$.\smallskip

\noindent{\it Remark}. As is evident from its construction the connection
$\Gamma$ on $\R^{2n}=\R^n_q\oplus\R^n_q$ is torsion free and non-metrical.
However, it does depend on the metric in the following way. The construction
(\ref{II.9})  of the $q,p$-components of the reflection map employs Euclidean
geodesics. The $q$ linearity of $q$-components and the $p$ linearity of the
$p$-components of $\sigma_x$ are responsible for the 0-blocks in the tensor
structure of $\Gamma$ and $R$.

\smallskip

  Stated covariantly formula (\ref{IV.4c}) reads
\begin{equation}\label{IV.7c}
\widetilde \Gamma_{ijk}= \frac 13 (\nabla_k^0 F_{ij} + \nabla_j^0  F_{ik})\,,
\end{equation}
where $\nabla^0_j$ are covariant derivatives on configuration space. In the
framework of the present paper $\nabla^0_j=\pr_j$ are just the Euclidean
derivatives on $\R^n_q$, but we can claim that formula (\ref{IV.7c}) actually
represents the magnetic contribution to the phase space connection also on
$T^*\M$ for any affine (in particular, Riemannian) configuration manifold $\M$
where $\nabla^0$ is non-trivial, for instance, non-flat. In the general case,
of course, the three other blocks in (\ref{IV.6n}) will no longer be zero
(compare with \cite{BNW98}). We postpone corresponding details to another
paper.
\smallskip

\begin{prop} The coefficients $c_2$ and $c_3$  in the $*$
product expansion {\rm(\ref{IV.1})} have the covariant form:
\begin{eqnarray} \label{c2}
c_2(\cdot,\cdot) &=& \langle \ola \nabla \Psi\ora
\nabla\rangle^2\,,\phantom{\Big)} \\
c_3(\cdot,\cdot) &=& \langle\ola\nabla\Psi\ora\nabla\rangle^3 +
R^{ijkl}(\ora\nabla_i\ora\nabla_j\ora\nabla_k\ola\nabla_l -
\ola\nabla_i\ola\nabla_j\ola\nabla_k\ora\nabla_l)\,. \label{c3}
\end{eqnarray}
Here $R^{ijkl} = R^i_{j'k'l'}\Psi^{j'j}\Psi^{k'k}\Psi^{l'l}$\,, and $\nabla$ is
the covariant derivative with respect to the symplectic connection $\Gamma$
{\rm(\ref{I.9})}.
\end{prop}

\pf  \ The non-covariant form of the $k=2$ coefficient reads
\begin{equation*}
c_2(f,g) = f\langle \ola D \Psi\ora D\rangle^2g + f M_0(\ola \pr, \ora \pr)
g\,.
\end{equation*} The $M_0$ factor above is
\begin{equation*}
M_0(u,v) = \frac 2 3 u^k u^s\pr_s F_{kl} v^l - \frac 2 3 u^k \pr_s F_{kl}v^l
v^s=- \Big(u^ku^s \wt \Gamma_{lsk}(q) v^l +u^k\wt \Gamma_{kls}(q)v^l v^s
\Big)\,.
\end{equation*}
Straightforward calculations show that
\begin{equation*}
\langle \ola D \Psi\ora D\rangle^2 +  M_0(\ola \pr, \ora \pr)  = \langle \ola
\nabla \Psi\ora \nabla\rangle^2 \,.
\end{equation*} This establishes (\ref{c2}).

Consider next the $O(\h^3)$ coefficient. It has the non-covariant form
\begin{equation*}
c_3(f,g) = f\Big( \langle \ola D \Psi\ora D\rangle^3 + 3  \langle \ola D
\Psi\ora D\rangle M_0(\ola \pr, \ora \pr) + 3M_1(\ola \pr, \ora \pr)\Big)g
\end{equation*} where
\begin{equation*}
M_1(u,v) = \frac 1 3 \Big(u^ku^su^r \pr^2_{sr} F_{kl} v^l + u^k \pr^2_{sr}
F_{kl} v^lv^sv^r\Big) = \wt R_{srkl}\Big(u^su^ru^kv^l - v^sv^rv^ku^l\Big)\,.
\end{equation*}  A tensor computation then shows that
\begin{equation*}
 \langle \ola D \Psi\ora D\rangle^3 + 3  \langle \ola D \Psi\ora D\rangle
M_0(\ola \pr, \ora \pr) + 2M_1(\ola \pr, \ora \pr) =  \langle \ola \nabla
\Psi\ora \nabla\rangle^3\,.
\end{equation*}
Thus the $O(\h^3)$ coefficient becomes
\begin{equation*}
c_3(f,g) = f\Big(\langle \ola \nabla \Psi\ora \nabla\rangle^3 + \wt
R_{srkl}(\ola\pr^s\ola\pr^r\ola\pr^k\ora\pr^l -
\ora\pr^s\ora\pr^r\ora\pr^k\ola\pr^l)\Big)g\,.
\end{equation*}
Finally, index raising by $\Psi(q)$ on $\wt R$ gives the covariant $c_3(f,g)$
expression in (\ref{c3}). $\square$

\section{Quantization with arbitrary electromagnetic fields}
\setcounter{equation}{0}

So far it has been assumed that the magnetic fields are static. Now we consider
the modifications in the Weyl quantization that arise when the electromagnetic
fields are time dependent. The manner in which the electric field enters the
symbol calculus  is made explicit.

First it is helpful to clarify the role of the vector potential $A$ in the
static Weyl quantization. Within the quantum phase space framework, the
potential $A$ never appears. The $\omega$ symplectic form and Poisson brackets,
the symplectic area $\Sigma$, the connection $\Gamma$, the $*$ product and its
expansion coefficients $c_k$ are all defined directly in terms of the magnetic
tensor $F$. However, the 2-point Valatin potential $A(q',q'')$ (\ref{III.10})
which is a non-local gauge invariant object, does appear spontaneously as a
contribution to the $l,r$ functions and is essential in the definition of the
reflection symmetry $\sigma_x$ and the groupoid product. Only when one goes to
the Hilbert space representation, via $f\rightarrow \f$,
\cf(\ref{I.4}),\,(\ref{II.1}), is any gauge fixing required. In order to
represent the quantizer $\Delta(x)$, a vector potential $A$ (consistent with
$F$) must be employed. One convenient gauge choice for $A$ is to use again the
Valatin potential with a fixed $2^{nd}$ argument. This was the option selected
in our prior work \cite{KO1}, where the fixed $2^{nd}$ argument was set to 0.

Let $B(t,q)$ denote the magnetic field. In the $n=3$, time dependent case the
2-form becomes
\begin{equation}\label{5a.1}
\omega(t) = dp\wedge dq + B_1(t,q)\,dq^2\wedge dq^3 + B_2(t,q)\,dq^3\wedge dq^1
+ B_3(t,q)\,dq^1\wedge dq^2\,.
\end{equation}
Similarly, the quantum commutation relations (\ref{I.2}) acquire time
dependence via the momentum components by $ [\p_j(t),\p_k(t)]=i\h
\epsilon_{jkl}B_l(t,\q)\,.$

Replacing the static $\omega$ with $\omega(t)$ in (\ref{I.7}) defines a $*$
product that is time dependent. This time dependence results from the $t$
varying  magnetic flux through the triangle $\delta(q_3,q_2,q_1)$. Likewise the
quantizer, the left, right coordinates, the refection symmetry $\sigma_x$ and
the magnetic connection all acquire an obvious $t$ dependence.

In order to fix the quantizer and obtain a unique irreducible Hilbert space
representation, the quantum coordinates (\ref{II.1}) need to be defined. Let us
work in the Coulomb gauge where the 4-vector potential
$a(t,q)\equiv\{-\phi(t,q),A(t,q)\}$ has a vanishing scalar component,
$\phi(t,q)=0$.  There is no loss of generality in this Coulomb gauge assumption
since given a general 4-vector one may, by a known unitary transformation,
always gauge away the scalar component. In the Coulomb gauge
\begin{equation}\label{5a.2}
B(t,q) = \nabla \times A(t,q)\,, \qquad E(t,q) = - \frac \pr{\pr t} A(t,q)\,.
\end{equation}

Define $\q,\p(t)$ by (\ref{II.1}) with the static $A(q)$ replaced by $A(t,q)$.
From (\ref{II.1}) and (\ref{5a.2}) it follows that
\begin{equation} \frac d{dt}\, \p(t) = E(t,\q)\,.
\end{equation}

This approach is based on the separation of time and space variables that is
needed for solving the Cauchy problem, see details in \cite{KO1}. With this
separation we observe that the quantum phase space coordinates have acquired
time dependent momentum components. The magnetic field $B(t,q)$ determines the
symplectic structure via (\ref{5a.1}), whereas in the Coulomb gauge the
electric field $E(t,q)$ generates the motion of the kinetic coordinates
$\p(t)$. The magnetic curvature $R^i_{jkl}(t,q)$ is time dependent on the phase
space $\R^6=\R^3_q\oplus\R^3_p$ and does not sense the electric field.

Another view point is to include the time and space variables together into the
configuration space $\R^4_{t,q}$. Then the symplectic form and the magnetic
curvature tensor on the space $\R^8=\R^4_{t,q}\oplus\R^4_{p_t,p}$ will now
depend on the
electric field as well.

\section{Zero magnetic curvature }
\setcounter{equation}{0} Let us return to the static situation. The magnetic
connection (\ref{I.9}) is determined by the first derivatives of the tensor
$F$.  Its curvature is determined by the second derivatives.

The simplest case is the {\it{homogeneous magnetic field}} that is $F= const$.
In this case the Christoffel symbols $\Gamma$ are just zero (in the Euclidean
basis), and the magnetic connection coincides with the Euclidean connection on
$\R^{2n} = T^*\R^n$.

    The second simple example is that of a {\it linear magnetic field}, that is
\begin{equation}\label{V.1}
\begin{array}{rl}
F_{ij}(q) &=F_{ij,k}\, q^k\,,\\[2ex]
F_{ij,k} = -F_{ji,k}\,,& \quad  F_{ij,k} + F_{jk,i} + F_{ki,j}=0 \,.
\end{array}
\end{equation}
This is the Lie algebra case; the commutation relations (\ref{I.2}) are linear.
In this case the magnetic connection becomes a constant,
\begin{equation}\label{V.2}
\wt\Gamma_{ijk} = {\frac 1 3} (F_{ij,k} + F_{ik,j})\,,
\end{equation}
and so the magnetic curvature is zero: $R=0$. The reflection symmetry
$\sigma_x$ in this case is realized by quadratic mappings
\begin{equation}\label{V.2a}
\sigma_x(y)= 2x-y - \left( \begin{array}{c} 0 \\ \wt
\Gamma_{\cdot,\,jk}(x_q-y_q)^j(x_q-y_q)^k
\end{array} \right)\,.
\end{equation}

Let us consider the following non-symplectic change of variables in $\R^{2n}$:
\begin{equation}\label{V.3}
q'= q\,, \qquad p' = p + A(q)\,.
\end{equation}
Here $A$ is a magnetic potential satisfying (\ref{II.2}). Under this
transformation the magnetic form $\omega$ is transformed into the canonical
form: $\omega' = dp'\w dq'$.

 For instance, one can take $A$ to be the Valatin potential with fixed a second argument $0$,
\begin{equation}\label{V.4}
A_i(q) = A_i(q,0) = {\frac 1 3} F_{ij,k}q^j q^k = {\frac 1 2} \wt\Gamma_{ijk}
q^jq^k\,
\end{equation}
which satisfies the radial gauge condition (\ref{III.9}), $ q^iA_i(q) = 0\,$.
For a quadratic $A$, the $x\rightarrow x'$ variable change maps the magnetic
connection $\Gamma$ into $\Gamma'=0$ (the Euclidean connection).

The form $\omega'$ and the connection $\Gamma'$ generate the usual Groenewold
$*$-product over $\R^{2n}$, which can be expressed both in the integral form
and in the derivative form
\begin{eqnarray}
\label{V.8} \big(f'*'g'\big)(x') &=& {\frac 1{(\pi\h)^{2n}}}\int \!\!\! \int
\exp \Big\{\ih \int_{M(x',y'z')} \omega' \Big\} f'(y')g'(z') dy'\,dz'\\[2ex]
\big(f'*'g'\big)(x') &=& f'(x')\exp\Big\{-{\frac{i\h}2} \langle\ola D' \Psi'
\ora D'\rangle\Big\} g'(x')\label{V.7} \,.
\end{eqnarray}
Here $\Psi'=[\begin{array}{cc}0&-I\\I&0\end{array}]$ is the Poisson tensor
corresponding to the symplectic form $\omega'$, the derivatives $D' = \pr/{\pr
x'}$ are taken with respect to the coordinates $x'=(q',p')$, and the membrane
$M(x',y',z')$ is just the triangle in $\R^{2n}$ with midpoints $x',y'z'$.

In formulas (\ref{V.7}), (\ref{V.8}) we denote the symbols $f',g'$ by prime
indices in order to emphasize that these functions are expressed in the new
coordinate system $x'=(q',p')$. Of course, there is a correspondence with the
functions in the previous (magnetic) coordinate system $x=(q,p)$, namely
\begin{equation*}\label{V.9}
f(x) =f'(x')\,, \qquad x\leftrightarrow x' \quad {\text {by}}\quad (\ref{V.3})
\end{equation*}
In the magnetic coordinates we have the magnetic product (\ref{I.7}). So the
question arises: does this magnetic product correspond to the Groenewold
product under this change of variables?

\begin{prop}\label{prop2}
Assume that the magnetic curvature $R$ is zero, that is the magnetic tensor $F$
is linear. Let the change of variables {\rm(\ref{V.3})} satisfy the radial
gauge condition, \ie the potential $A$ is given by {\rm(\ref{V.4})}. Then under
this quadratic change of variables $x\leftrightarrow x'$ the magnetic product
{\rm(\ref{I.7})} generated by the form $\omega$ {\rm(\ref{I.1})} corresponds to
the Groenewold product {\rm(\ref{V.8})} generated by the form $\omega' = dp'\w
dq'$.

In particular, the Groenewold differential formula {\rm(\ref{V.7})} implies the
following representation of the magnetic product,
\begin{equation}\label{V.10}
(f*g)(x) = f(x)\exp\Big\{-{\frac {i\h}2} \ola \nabla \Psi(x) \ora \nabla\Big\}
g(x)\,,
\end{equation}
where $\Psi$ is the magnetic Poisson tensor, and where the covariant
derivatives are given by the flat magnetic connection {\rm(\ref{V.2})}.
\end{prop}
\pf \, First perform the variable change $x \rightarrow x'$ in the M-integral
representation of the $*$ product, \cf (\ref{III.2}), (\ref{III.3}). If $A$ is
quadratic, one readily finds that
 \begin{equation*}
\int\!\!\int\exp\Big\{ \frac i \hbar \int_{M(z,y,x)}\omega \Big\} f(y) g(x)\,dy
\,d x = \int\!\!\int\exp\Big\{ \frac i \hbar \int_{M(z'(z),y',x')}\omega'
\Big\} f'(y') g'(x')\,dy' \,d x'\,.
\end{equation*}
This establishes that $\big(f*g\big)(x)=\big(f'*'g'\big)(x'(x))$ and verifies
that $f*g$ is given by the right hand side of (\ref{V.7}). Now implement the
inverse transform $x'\rightarrow x$ and employ $f'\langle\ola D' \Psi'\ora
D'\rangle^N g' = f\langle\ola \nabla \Psi(x) \ora \nabla\rangle^N g$ to obtain
(\ref{V.10}). $\square$

The content of Proposition~\ref{prop2} agrees well with known formulas for
formal $*$ products over flat symplectic manifolds \cite{BBF78}, \cite{BT90}.
But we see that our formula (\ref{V.10}) actually holds for the non-formal
strict $*$ product (\ref{I.7}) which has the operator representation
(\ref{I.4}), and that the connection $\nabla$ in (\ref{V.10}) is exactly the
magnetic connection (\ref{I.9})

The change of variables (\ref{V.3}) can also be carried out for general
non-linear tensors $F$.  Again the Groenewold formula (\ref{V.7}) generates in
this way a certain product
\begin{equation}\label{V.11}
(f\times g)(x) \equiv f(x)\exp\Big\{-{\frac {i\h}2} \langle\ola \nabla^A
\Psi(x) \ora \nabla^A\rangle\Big\} g(x)\,,
\end{equation} where $\nabla^A$ corresponds to the flat connection with
Christoffel symbol components $\wt \Gamma_{ijk} = \pr^2_{jk}A_i(q)$. However,
here the approach of deriving the the magnetic $*$ product through the variable
change (\ref{V.3}) fails. The product (\ref{V.11}) is not related to the
magnetic product and the flat connection $\nabla^A$ is not the magnetic
connection, if $F$ is not linear.

\section{Constant magnetic curvature }
\setcounter{equation}{0} Constant magnetic curvature means that the tensor $F$
is quadratic. There is no loss in generality here in assuming that $F$ is
purely quadratic with no linear component. In detail
\begin{equation}\label{VI.1}
\begin{array}{lll}
\qquad F_{ij}(q) =F_{ij,kl}\, q^k q^l\,,& F_{ij,kl} = -F_{ji,kl}\,,& \quad
F_{ij,kl} = F_{ij,lk}\,,\qquad
\\[2ex]\qquad
F_{ij,kl} + F_{jk,il} +F_{ki,jl} =0\,. & &
\end{array}
\end{equation}
In this case the commutation relations (\ref{I.2}) are quadratic and the
algebra is not a Lie algebra.

The magnetic connection and curvature are given by
\begin{equation}\label{VI.2}
\wt\Gamma_{ijk}(q) = {\frac 2 3}\big(F_{ij,kl} + F_{ik,jl})\,q^l\,, \qquad \wt
R_{ijkl} = {\frac 2 3}F_{kl,ij}\,.
\end{equation}
The magnetic reflection is still a quadratic mapping:
\begin{equation}\label{VI.2a}
\sigma_x(y)= 2x-y - \left( \begin{array}{c} 0 \\
\wt \Gamma(x_q)_{\cdot,jk}(x_q-y_q)^j(x_q-y_q)^k
\end{array} \right)\,.
\end{equation}

\begin{lem}\label{lem6}
Let the Faraday $F$ tensor be quadratic, that is, the magnetic curvature be
constant. Then the reflections $\sigma_x$ {\rm(\ref{VI.2a})}

\itn{i} are affine with respect to the magnetic connection (map geodesics into
geodesics);

\itn{ii} satisfy the symmetry condition {\rm(\ref{I.11})};

\itn{iii}  coincide with geodesic reflections generated by the magnetic
connection.
\end{lem}

\pf \,\itn{i} Let $\gamma(\xi)=(q(\xi),p(\xi))\,,\, \xi\in[-1,1]$ be a generic
magnetic geodesic. The zero block structure of $\Gamma$ allows one to state the
geodesic equation of motion as
\begin{equation}\label{7.4}
\ddot{\gamma}^i(\xi) + \Gamma^i_{jk}(q(\xi))\,\dot q^j(\xi)\dot q^k(\xi)= 0\,,
\qquad j,k\leq n\,.
\end{equation}
Since $\Gamma^i_{jk}=0$ for $i\leq n, \quad \ddot{q}(\xi)=0\,$ and
$\,\dot{q}(\xi)={\rm const}$.

Set $\gamma'(\xi)\equiv \sigma_x(\gamma(\xi))$; we must show $\gamma'$ is a
geodesic. For $\sigma_x$'s given by (\ref{VI.2a}), the second derivative of
$\gamma'$ is
\begin{equation*}\label{VI.3a}
\ddot{\gamma}'^i(\xi) = -\ddot{\gamma}^i(\xi) -2\Gamma^i_{jk}(x_q)\dot q^j\dot
q^k\,.
\end{equation*}
Use $q(\xi) + q'(\xi) = 2x_q$; the $q$-linearity of $\Gamma$, $2
\Gamma^i_{jk}(x_q) = \Gamma^i_{jk}(q(\xi)) + \Gamma^i_{jk}(q'(\xi))$, and
$\dot{q}= -\dot{q'}$ to show the above identity is equivalent to
\begin{equation}\label{VI.3b}
\ddot{\gamma}'^i(\xi) + \Gamma^i_{jk}(q'(\xi))\,\dot q'^j\dot
q'^k=-\Big(\ddot{\gamma}^i(\xi) + \Gamma^i_{jk}(q(\xi))\,\dot q^j\dot q^k\Big)
=0\,.
\end{equation} Thus $\gamma'$ is a geodesic.

A similar argument verifies $\itn{ii}$; item $\itn{iii}$ results from a
straight forward algebraic calculation. $\square$

Note that property $\itn{ii}$ means that in the quadratic case the reflections
$\sigma_x$ {\rm(\ref{VI.2a})} determine the symmetric symplectic structure on
the phase space $\R^{2n}$ in the sense of \cite{BCG95}.

\begin{cor}\label{cor16}
If the magnetic tensor is quadratic, then the magnetic product can be
represented by formula {\rm(\ref{I.7})} using membranes $\Sigma(z,y,x)$ bounded
by three magnetic geodesics with midpoints $z,y,x$.
\end{cor}

Considering higher terms in the formal $\hbar$-power expansion (\ref{IV.1}) in
the covariant form (\ref{I.8}), (\ref{c2}), and (\ref{c3}) we can conjecture
the following generalization of the Groenewold representation.

\begin{conjecture}
If the magnetic curvature is constant (that is, $F$ is quadratic), then
\begin{equation}\label{VI.3}
f*g = f\exp\Big\{-{\frac {i\h}2} \langle\ola \nabla \Psi(x) \ora \nabla\rangle\
+ \frac{i\h^3}{48} R^{ijkl}(\ora\nabla_i\ora\nabla_j\ora\nabla_k\ola\nabla_l -
\ola\nabla_i\ola\nabla_j\ola\nabla_k\ora\nabla_l)\Big\} g \,,
\end{equation}
where $R$ is the magnetic curvature with raised indices, and $\nabla$ is the
magnetic connection.
\end{conjecture}

\section{Conclusions}
\setcounter{equation}{0}

The quantum coordinates $\q^j$ and kinetic momenta $\p_k$ of a charged particle
know about the presence (or absence) of magnetic field via the commutation
relations between momenta \cite{Dy90}. In general, these commutation relations
are non-linear.

Weyl-symmetrized functions in the operators  $\q,\p$  form an algebra. The
symbol image of the multiplication in this magnetic algebra can be represented
(exactly) by a  simple integral formula (\ref{I.7}) via the magnetic symplectic
form $\omega$ and membranes $\Sigma$ having a groupoid-consistent boundary.
This is an example of perfectly quantizable phase space.

The groupoid structure, corresponding to the form $\omega$, is controlled by a
family of magnetic reflections (\ref{II.9}), which are generated by the regular
left and right representations (\ref{III.7}) of the magnetic algebra.

The family of magnetic reflections determines a symplectic connection $\Gamma$,
(\ref{I.9}). The magnetic $*$ product (\ref{I.7}) has a covariant derivative
asymptotic expansion (\ref{I.8}) whose $\h^3$-term (\ref{I.11}) is given by the
curvature of this connection.

In zero curvature case, the magnetic $*$ product can be independently recovered
from the standard (non-magnetic) Groenewold exponential formula by a
non-symplectic change of variables. The resultant exponential formula is stated
in terms of covariant derivatives generated by the magnetic connection.

The case of constant (but non-zero) curvature represents an interesting example
of a symplectic symmetric space. The magnetic $*$ product in this case is given
either via a geodesic bounded membrane area or via the explicit covariant
differential formula (\ref{VI.3}) with magnetic curvature tensor in the
exponent.

The $\h^3$ term in the $\h\rightarrow 0$ expansion of our magnetic $*$ product,
in the constant curvature case, is different from the corresponding term in the
known Bieliavsky--Cahen--Gutt product \cite{BCG95} (given for general
symplectic symmetric spaces). In the general non-constant curvature case a
difference of numerical coefficient in the $\h^3$ terms can also be observed in
the comparison with the Fedosov deformation expansion \cite{BF96}.

Of course, on the level of formal $\h$ power series all associative $*$
products are equivalent \cite{Kon97}. But only some of them, like our
expansions (\ref{I.8}), (\ref{I.10}), are related to exact products possessing
an irreducible operator representation in a Hilbert space. Such an operator
representation does not allow  generic $\h$-pseudodifferential transformations
(allowed by the formal $*$ products).  The condition that the $*$ product admit
an exact irreducible operator representation in essence restricts the variety
of $*$ products (see in \cite{Rief89}).

The magnetic connection and its curvature, which we extract from from the
quantum algebra, is determined by the magnetic tensor $F$, but not in the way
as it usually appears in gauge field theory via $U(1)$-line bundles
\cite{BS82}, nor in the way suggested by Weyl nearly a century ago
\cite{Weyl18}. For instance, the magnetic connection is defined on the phase
space rather than on the configuration space. This magnetic connection is
clearly important in the quantization process, but we also anticipate that it
can be observed in the dynamical and spectral problems involving magnetic
fields such as the Fock-Landau level problem in the inhomogeneous case.

\bigskip
\bigskip {\bf Acknowledgments}. The authors wish to thank S.~A.~Fulling for suggesting various improvements to the paper.  The first
author is grateful to Russian Basic Research Foundation for partial support
(grant No.~02-01-00952). The research of T.A.O. is supported by a grant from
Natural Sciences and Engineering Research Council of Canada. The authors thank
the Winnipeg Institute of Theoretical Physics for its continuing support.

\renewcommand{\theequation}{\Alph{section}.\arabic{equation}}


\begin{thebibliography}{10}

\bibitem{BBF78}
{F.~Bayen, M.~Flato, C.~Fronsdal, A.~Lichnerowicz, and D.~Sternheimer}.
\newblock Deformation theory and quantization.
\newblock {\em Ann. Phys. (N.Y.)}, 111:61--110, 1978.

\bibitem{Raw91}
{J.~Rawnsley}.
\newblock Deformation quantization of {K}\"{a}hler manifolds.
\newblock In P.~Donato et~al, editor, {\em In Symplectic Geometry and
  Mathematical Physics}, pages 366--377, Basel-Boston, 1991. Birkhauser.

\bibitem{OMY91}
{H.~Omori, Y. Maeda and A. Yoskioka}.
\newblock Weyl manifolds and deformation quantization.
\newblock {\em Advances in Mathematics}, 85:224--255, 1991.

\bibitem{BF96}
{B. Fedosov}.
\newblock {\em Deformation Quantization}.
\newblock Akademie Verlag, Berlin, 1996.

\bibitem{Kon97}
{M. Kontsevich}.
\newblock Deformation quantization of {P}oisson manifolds, {I}.
\newblock {\em arXiv:~math, q-alg/9709040}, 1997.

\bibitem{BNW98}
{M.~Bordemann, N.~Neumaier, S.~Waldmann}.
\newblock Homogeneous {F}edosov star products on cotangent bundles~{I}: Weyl
  and standard ordering with differential operator representation.
\newblock {\em Commun. Math. Phys.}, 198:363--396, 1998.

\bibitem{Rief89}
M.~A. Rieffel.
\newblock Deformation quantization for actions of ${R}^d$.
\newblock In {\em Memoirs AMS}, volume 106, pages 1--93, 1993.

\bibitem{Biel01}
{P.~Bieliavsky}.
\newblock Strict quantization of solvable symmetric spaces.
\newblock {\em J.~Sympl. Geom.}, 1:269--320, 2002.

\bibitem{Lan93}
N.~P. Landsman.
\newblock Strict deformation quantization of a particle in external
  gravitational and {Y}ang-{M}ills fields.
\newblock {\em J. Geom. Phys.}, 12:93--132, 1993.

\bibitem{Car99}
{J. F. Cari\~nena, J. Clemente-Gallardo, E. Follana, J. M. Gracia-Bond\'ia, A.
  Rivero and J. C. Varilly}.
\newblock Connes' tangent groupoid and strict quantization.
\newblock {\em J. Geom. Phys.}, 32:79--96, 1999.

\bibitem{FAB72}
F.~A. Berezin.
\newblock Quantization.
\newblock {\em Math. USSR-Izv.}, 8:1109--1165, 1974.

\bibitem{BT90}
{I.~A.~Batalin and I.~V.~Tyutin}.
\newblock Quantum geometry of symbols and operators.
\newblock {\em Nuc. Phys.}, B345:645--658, 1990.

\bibitem{GB92}
{J.~M.~Gracia-Bondia}.
\newblock Generalized {M}oyal quantization on homogenous symplectic spaces.
\newblock {\em Contemporary Mathematics}, 134:93--114, 1992.

\bibitem{BS82}
B.~Schutz.
\newblock {\em Geometrical Methods of Mathematical Physics}.
\newblock Cambridge University Press, Cambridge, 1982.

\bibitem{Mas73}
{V.~P.~Maslov}.
\newblock {\em Operational {M}ethods}.
\newblock Nauka, Moscow, 1973. English transl.: MIR, 1976.

\bibitem{STR57}
R.~L. Stratonovich.
\newblock On distributions in representation space.
\newblock {\em Sov. Phys. JETP}, 4:891--898, 1957.

\bibitem{VGB89}
{J. C. Varilly and J. M. Garcia--Bondia}.
\newblock The {M}oyal representation of spin.
\newblock {\em Ann. Phys. (N.Y.)}, 190:107--148, 1989.

\bibitem{Fock}
{V.~A.~Fock}.
\newblock On the canonical transformation in classical and quantum mechanics.
\newblock {\em Vestnik Leningrad State Univ.}, 16:67--71, 1959 (in Russian).

\bibitem{KO1}
{M. V. Karasev and T. A. Osborn}.
\newblock Symplectic areas, quantization, and dynamics in electromagnetic
  fields.
\newblock {\em J. Math. Phys.}, 43:756--788, 2002 and arXiv: phys., quant-ph/0002041, 2000.

\bibitem{KM91}
M.~V. Karasev and V.~P. Maslov.
\newblock {\em Nonlinear Poisson Brackets. Geometry and Quantization}, Nauka, Moscow, 1991; English transl. in  volume
  119 of {\em Ser. Translations of Mathematical Monographs}.
\newblock AMS, Providence, 1993.

\bibitem{KO02}
{M. V. Karasev and T. A. Osborn}.
\newblock Magnetic curvature of quantum phase space.
\newblock In M.~A.~Vasiliev A.~M.~Semikhatov and V.~N. Zaikin, editors, {\em
  Third International Sakharov Conference on Physics}, volume~1, pages
  153--162, Moscow, 2003. Scientific World.

\bibitem{AW94}
A.~Weinstein.
\newblock Traces and triangles in symmetric symplectic spaces.
\newblock {\em Contemp. Math.}, 179:262--27, 1994.

\bibitem{BCG95}
{P. Bieliavsky, M. Cahen, and S. Gutt}.
\newblock Symmetric symplectic manifolds and deformation quantization.
\newblock In J.~Bertrand et~al, editor, {\em Modern group methods in
  {P}hysics}, volume~18 of {\em Mathematical physics studies}, pages 63--73,
  Paris, 1995. Kluwer.

\bibitem{Loo69}
{O.~Loos}.
\newblock {\em Symmetric {S}paces}, volume~1.
\newblock W. A. Benjamin, New York, 1969.

\bibitem{CART26}
{\'E.~Cartan}.
\newblock Sur une classe remarquable d'espaces de {R}iemann.
\newblock {\em Bull. Soc. Math. France}, 54:214--264, 1926.

\bibitem{ANDER69}
{R.~F.~V.~Anderson}.
\newblock On the {W}eyl functional calculus.
\newblock {\em J. Funct. Anal.}, 4:240--267, 1969.

\bibitem{ANDER70}
{R.~F.~V.~Anderson}.
\newblock On the {W}eyl functional calculus.
\newblock {\em J. Funct. Anal.}, 6:110--115, 1970.

\bibitem{Str56}
R.~L. Stratonovich.
\newblock A gauge invariant analog of the {W}igner distribution.
\newblock {\em Sov. Phys. D}, 1:414--418, 1956.

\bibitem{Gro76}
A.~Grossmann.
\newblock Parity operator and quantization of $\delta$-functions.
\newblock {\em Commun. Math. Phys.}, 48:191--194, 1976.

\bibitem{Roy77}
A.~Royer.
\newblock Wigner function as the expectation value of a parity operator.
\newblock {\em Phys. Rev. A}, 15:449--450, 1977.

\bibitem{Ber77}
M.~V. Berry.
\newblock Semi-classical mechanics in phase space: a study of {W}igner's
  function.
\newblock {\em Phil. Trans. R. Soc. Lond. A}, 287:237--271, 1977.

\bibitem{Mar79}
M.~S. Marinov.
\newblock An alternative to the {H}amilton--{J}acobi approach in classical
  mechanics.
\newblock {\em J. Phys. A: Math. Gen}, 12:31--47, 1979.

\bibitem{DMO92}
{V.~V.~Dodonov, V.~I.~Man'ko and D.~L.~Ossipov}.
\newblock Gradient-invariant {W}eyl representation. {A}n oscillator in an
  inhomogeneous magnetic field.
\newblock In M.~A. Markov, editor, {\em Theory of Nonstationary Quantum
  Oscillators}, volume 198 of {\em Proceedings of the Lebedev Physical
  Institute}, pages 1--58, 1992.

\bibitem{Val54}
J.~G. Valatin.
\newblock Singularities of electron kernel functions in an external
  electromagnetic field.
\newblock {\em Proc. Soc. A}, 222:93--108, 1954.

\bibitem{Foll89}
G.~B. Folland.
\newblock {\em Harmonic analysis in Phase Space}.
\newblock Princeton University Press, Princeton, 1989.

\bibitem{Hor79}
L.~Hormander.
\newblock The {W}eyl calculus of pseudo-differential operators.
\newblock {\em Comm. Pure Appl. Math.}, 32:359--443, 1979.

\bibitem{KN78}
{M. V. Karasev and V. E. Nazaikinskii}.
\newblock On quantization of rapidly oscillating symbols.
\newblock {\em Math. USSR-Sb.}, 34:737--764, 1978.

\bibitem{Mul99}
M.~M{\"u}ller.
\newblock Product rule for gauge invariant {W}eyl symbols and its application
  to the semiclassical description of guiding centre motion.
\newblock {\em J. Phys. A}, 32:1035--1052, 1999.

\bibitem{Dy90}
{F.~J.~Dyson}.
\newblock Feynman's proof of the {M}axwell equations.
\newblock {\em Amer. J. Physics}, 58:209--211, 1990.

\bibitem{Weyl18}
{H. Weyl}.
\newblock {G}ravitation and {E}lectricity.
\newblock {\em Sitzungsber. Preuss. Akad. Berlin}, page 465, 1918.

\end{thebibliography}
\end{document}